\documentclass[12pt]{article}
\usepackage{epsfig}

\def\be{\begin{equation}}
\def\ee{\end{equation}}
\def\ba{\begin{array}{c}}
\def\ea{\end{array}}

\def\ben{$$}
\def\een{$$}
\newcommand{\bbr}{\br\!\br}
\newcommand{\kkt}{\kt\!\kt}

\newcommand{\kt}{\rangle}
\newcommand{\br}{\langle}
\begin{document}

\titlepage

\vspace{.35cm}

 \begin{center}{\Large \bf

The complete menu of eligible metrics for a family of toy
Hamiltonians $H \neq H^\dagger$ with real spectra.

  }\end{center}

\vspace{10mm}

 \begin{center}

 {\bf Miloslav Znojil}

 \vspace{3mm}
Nuclear Physics Institute ASCR,

250 68 \v{R}e\v{z}, Czech Republic

{e-mail: znojil@ujf.cas.cz}

\vspace{3mm}


\end{center}

\vspace{5mm}


\section*{Abstract}

An elementary set of non-Hermitian $N$ by $N$ matrices
$H^{(N)}(\lambda) \neq \left [ H^{(N)}(\lambda)\right ]^\dagger$
with real spectra is considered, assuming that each of these
matrices represents a selfadjoint quantum Hamiltonian in an {\it
ad hoc} Hilbert space of states ${\cal H}^{(physical)}$. The
problem of an explicit specification of all of these spaces (i.e.,
in essence, of all of the eligible  {\it ad hoc} inner products
and metric operators $\Theta$) is addressed. The problem is shown
exactly solvable and, for every size $N=2,4,\ldots$ and parameter
$\lambda \in (-1,1)$ in matrix $H^{(N)}(\lambda)$, the complete
$N-$parametric set of metrics
$\Theta^{(N)}_{\alpha_1,\alpha_2,\ldots,\alpha_N}(\lambda)$ is
recurrently defined by closed formula.

\newpage

\section{Introduction \label{I} }

\subsection{The concept of the metric $\Theta$ in Hilbert space
${\cal H}^{(physical)}$\label{Ia}}

One-dimensional quantum systems described, in units $\hbar = 2m =
1$, by the ordinary differential Schr\"{o}dinger equation
 \be
 H\,\psi(x)=E\,\psi(x)\,,
 \label{basic}
 \ \ \ \ \ \ \ \
 H= -\frac{d^2}{dx^2} + V(x)
 \,
 \label{basicss}
 \ee
serve as a universal testing ground for the ideas, methods and
techniques of quantum mechanics. One works with the standard
representation $L^2(I\!\!R)$ of the Hilbert space of states where
the bound states are normalized in usual manner and where the
Hamiltonian itself is self-adjoint,
 \be
 \int_{I\!\!R}\,\psi^*(x)\,\psi(x)\,dx
 = 1\,,\ \ \ \ \ \
   H = H^\dagger
 \,.
 \label{usuall}
 \ee
In parallel, the scattering solutions $\psi(x)$ of
eq.~(\ref{basic}) offer the simplest illustration of the
delocalized waves which must remain compatible with the unitarity
of the time evolution, etc. (cf., e.g., ref. \cite{Fluegge} for
numerous illustrations).

The transparency of such an elementary implementation of quantum
theory can prove deceptive. People often forget that the
requirement of the Hermiticity of $H$ in $L^2(I\!\!R)$ can be
replaced by an alternative, equally acceptable requirement $H =
H^\ddagger$ of the Hermiticity of the {same} operator in {another}
Hilbert space ${\cal H}^{(physical)} \ \neq\ L^2(I\!\!R)$ where a
{different} definition of the inner product $( \psi,\psi')_\Theta$
is employed,
 \be
   ( \psi,\psi')_\Theta = \int_{I\!\!R^2}\,\psi^*(x)\,
 \Theta(x,x')\,\psi'(x')\,dx\,dx'\,,\ \ \ \ \ \
   H = H^\ddagger\ \equiv\ \Theta^{-1}H^\dagger\,\Theta
 \,.
 \label{newh}
 \ee
Although the space $L^2(I\!\!R)$ remains unchanged as a vector
space, the definition of the ``correct" linear functionals (i.e.,
the definition of the mapping ${\cal T}$ of the ``ket-vectors"
upon ``bra-vectors") is less elementary in ${\cal
H}^{(physical)}$,
 \be
 {\cal T}\,\psi(x) = \int_{I\!\!R}\,\psi^*(x')\,
 \Theta(x',x)\,dx'\,.
 \label{newhde}
 \ee
Misunderstandings may emerge whenever the metric is nontrivial,
$\Theta =\Theta^\dagger\neq I$. Then,  our Hamiltonian $H=
H^\ddagger$ appears non-Hermitian in the conventional Hilbert
space ${\cal H}^{(Dirac)} = L^2(I\!\!R)$. The paradox has an
elementary resolution since the space ${\cal H}^{(Dirac)}$ with
its standard inner product is {\em not} the space of states of the
system in question (cf. Appendix A for a brief recollection of a
few concrete illustrative examples with this ``cryptohermiticity"
\cite{Smilga} property).

\subsection{The\label{Ib} problem of the ambiguity of the metric $\Theta=\Theta(H)$ }

The deeper study of the similarity (\ref{newh}) between an
operator $H$ and its adjoint $H^\dagger$ in a preselected space
${\cal H}^{(Dirac)}$ dates back to the early sixties
\cite{Dieudonne}. In physics, the first use of such a feature,
i.e., of the so called quasi-Hermiticity constraint
 \be
   \Theta\,H =  H^\dagger\,\Theta\,\label{tarov}
 \label{htot}
 \ee
imposed upon a sufficiently nontrivial and realistic Hamiltonian
$H\neq H^\dagger$ emerged much later \cite{Geyer}. Still, up to
now the subject remains full of open questions. One of the most
challenging ones concerns the ambiguity of the metric $\Theta$
assigned, via eq.~(\ref{htot}), to a given Hamiltonian operator
$H$. Indeed, eq.~(\ref{htot}) itself defines ``too many"
alternative physical metric operators $\Theta=\Theta(H)$. An
explicit constructive illustration of such an ambiguity of the
assignment $H \to \Theta$ given in section 5 of ref. \cite{cubic}
employs a {\em free} Hamiltonian $H=H_0$ which is complemented by
a {two-parametric} family of non-trivial metric operators
 \be
 \Theta_0^{(Mostafazadeh)}(F,K)=e^{-F}
 ({\rm cosh} K -  {\cal P}{\rm sinh} K)
 \label{most}
 \ee
where ${\cal P}$ denotes parity and where $F$ and $K$ are
arbitrary real numbers.

The first discussion of the general problem of the ambiguity of
$\Theta(H)$ has been published by Scholtz et al \cite{Geyer}. They
emphasized that besides the Hamiltonian $H$ itself, {any other}
operator ${\cal O}={\cal O}_j$ of an observable quantity in ${\cal
H}^{(physical)}$ must obey the {same} Hermiticity relation as $H\
\equiv\ {\cal O}_0$. In an opposite direction, {any} eligible
{physical} metric operator $\Theta$ must remain compatible with
the corresponding set of the quasi-Hermiticity relations
 \be
   \Theta\,{\cal O}_j =  {\cal O}^\dagger_j\,\Theta\,,\ \ \ \ \ \
 j=(0), 1, 2, \ldots\,.
 \label{newhtot}
 \ee
These requirements reduce the ambiguity in $\Theta$ at every index
$j$. In this sense, the choice of the physical metric
$\Theta=\Theta(H)$ can be made, in principle, unique.

We intend to return to the problem of the ambiguity of the metric
in what follows. One of our reasons is that the  universal
strategy represented by requirement (\ref{newhtot}) is in fact
rarely successful in practice. The solution of the {\em complete}
set of the linear operator relations (\ref{newhtot}) appears to be
hardly feasible. Typically, just $j=0$ is considered and a {\em
particular } solution $\Theta(H)$ of eq. (\ref{htot}) is sought
for. Appendix B reviews a few alternative proposals of making the
Hamiltonian-dependent metric operator $\Theta(H)$ unique in such a
case.

\subsection{\label{Ic}Unique metrics in certain matrix models of scattering }



A schematic comparison of a few alternative techniques of the
removal of the ambiguity of metrics $\Theta(H)\neq I$ has been
performed in our two brief comments \cite{plb}. We restricted our
attention to the mere two-dimensional Hilbert spaces ${\cal
H}^{(physical)}$. Using an elementary set of two-by-two matrices
$H=H^{(2)}$ we compared the merits and shortcomings of various
versions of non-equivalent $\Theta$s. Due to the simplicity of the
space we were able to base our analysis on an {explicit}
construction of {\em all} the solutions $\Theta^{(2)}$ of eq.
(\ref{htot}).

In some sense we shall just extend such a study to certain less
trivial Hamiltonian matrices $H=H^{(N)}$ in what follows. The
practical feasibility of such a project relies on suitable
simplifications. It is obvious that for a general matrix $H^{(N)}$
one could hardly consider its size in the range $N>4$
\cite{determ}.

One of the most natural, anharmonic-oscillator-inspired choices of
the simplified tridiagonal matrices $H^{(N)}$ was proposed in
ref.~\cite{minimal}. Even these models with the number of variable
matrix elements limited to $N/2$ appeared to be only tractable
numerically \cite{minimalII}. In ref.~\cite{ojonesovi} this
observation led us to the most drastic reduction of the allowed
number $k$ of the variable matrix elements in $H^{(N)}(\lambda_1,
\ldots,\lambda_k)$ to the smallest integers $k = 1, 2$ and $3$.

It was a pleasant surprise to discover that the latter choice
proved extraordinarily successful. Without difficulties we were
able to consider all the matrix sizes $N=2K$ including even the
limiting case of $N = \infty$.  The ``exceptional" choice of
$N=\infty$ made us ready to study and solve certain difficult
conceptual problems in scattering theory (cf. the text of
ref.~\cite{ojonesovi} for more details). One of the reasons was
that already the one-parametric matrix model of dynamics
 \be
 H=H(\lambda)=
 \left [\begin {array}{rrrc|crrr}
  \ddots&\ddots&&&&&&
  \\{}
  \ddots&2&-1&&&&&
 \\{}
 &-1&2&-1&&&&
 \\{}
 &&-1&2&-1-\lambda&&&
 \\
  \hline
 &&&  -1+\lambda&2&-1&&
 \\{}
 &&&&-1&2&-1&
 \\{}
 &&&&&-1&2&\ddots
 \\{}
 &&& &&&\ddots&\ddots
 \end {array}\right ]\,
 \label{tridif} \label{hihi}
 \ee
proved compatible with the  diagonal matrix solution of
eq.~(\ref{htot}),
 \be
 \Theta[H(\lambda)]=
 \left [\begin {array}{rrrc|crrr}
  \ddots&&&&&&&
  \\{}
  &1-\lambda&&&&&&
 \\{}
 &&1-\lambda&&&&&
 \\{}
 &&&1-\lambda&&&&
 \\
  \hline
 &&&&1+\lambda&&&
 \\{}
 &&&&&1+\lambda&&
 \\{}
 &&&&&&1+\lambda&
 \\{}
 &&& &&&&\ddots
 \end {array}\right ]\,.
 \label{metr}
 \ee
The existence of such a local metric was already considered
improbable in the dedicated phenomenological literature
\cite{Jones} and we felt encouraged us to generalize the explicit
formula (\ref{metr}) to Hamiltonians with more parameters (cf.
\cite{ojonesovi}) and to some matrices with different structure
(cf. \cite{jpaft}).

Here, due to the lack of space, we shall skip all the similar
enhancements of sophistication. Rather, we shall return to the
bound-state problems where $N<\infty$ in what follows. Before
doing so we should add a remark that there exists an amazingly
close and direct connection between the two apparently independent
sample Hamiltonians given by eqs.~(\ref{basic}) and (\ref{hihi}).
In Appendices C ad D we shall explain this relationship in more
detail.

\subsection{\label{Icd}Matrix models and bound states }
%
%
%

In our present return to model (\ref{hihi}) and to the bound
states we shall consider all the truncated, finite-dimensional
matrix descendants $H^{(N)}$ of eq.~(\ref{tridif}) with
truncations $N=2K$. For this sequence of one-parametric toy
Hamiltonians
 \be
 H^{(2)}(\lambda)=
 \left [\begin {array}{cc} 2&-1-\lambda
 \\{}-1+\lambda&2\end {array}
 \right ]\,,
 \label{dvojka}
 \ee
 \be
 H^{(4)}(\lambda)=
 \left [\begin {array}{cccc} 2&-1&0&0\\{}-1&2&-1-\lambda&0
\\{}0&-1+\lambda&2&-1\\{}0&0&-1&2
\end {array}\right ]\,,
\label{cetyrki}
 \ee
 \be
 H^{(6)}(\lambda)=
 \left [\begin {array}{cccccc}
  2&-1&0&0&0&0
 \\{}-1&2&-1&0&0&0
 \\{}0&-1&2&-1-\lambda&0&0
 \\{}0&0&-1+\lambda&2&-1&0
 \\{}0&0&0&-1&2&-1
 \\{}0&0&0&0&-1&2
\end {array}\right ]\,
\label{cetyrkibe}
 \ee
etc the main result of our present paper will be the explicit
construction of the respective {\em complete} sets of {\em all}
the metrics $\Theta^{(N)}(\lambda)$ in closed form.

Such a project is nontrivial since the bound-state wave functions
remain localized so that there is no point in demanding the
asymptotic locality constraint which made the scattering metric
virtually unique in \cite{ojonesovi}. One can arrive at a really
satisfactory physical interpretation of the bound-state system
only via an {\em exhaustive} knowledge of {\em all} the metrics
$\Theta(H)$ allowed by eq.~(\ref{htot}).

In an introductory part of our present paper we shall set
$\lambda=0$ in $H^{(N)}(\lambda)$. In section \ref{method} this
simplification will help us to explain our method of construction
of all the admissible metrics $\Theta^{(N)}(0)=\Theta^{(N)}_0$. In
essence, we shall combine the brute-force symbolic-manipulation
constructions performed at the first few $N=2,4,\ldots$ with the
subsequent extrapolation of the resulting closed formulae towards
all the even integers $N=2K$.

In the second half of our paper (cf. sections \ref{Vab} and
\ref{VIab}) we shall return to the nontrivial, asymmetric
Hamiltonian matrices $H=H^{(N)}(\lambda)$ with $\lambda \neq 0$.
Firstly, in section \ref{Vab} we shall solve eq.~(\ref{htot}) for
the first three models (\ref{dvojka}), (\ref{cetyrki}) and
(\ref{cetyrkibe}). In section \ref{VIab} we shall then extrapolate
the resulting triplet of metrics to all the superscripts $N=2K$.
The closed formula for {all} of the solutions
$\Theta^{(N)}(\lambda)$ of eq.~(\ref{htot}) will be obtained as
our main result.

In section \ref{summary} we shall summarize our message while
Appendices A - D will complement it by a few additional remarks
and technical notes.

\section{The description of the method: $\lambda=0$ \label{method}}

\subsection{Trivial starting point: All the metrics at $N=2$.
\label{mduob} }

In the light of ref.~\cite{plb} the simplest possible
two-dimensional Hamiltonian
 \ben
 H_0=\left [\begin {array}{rr} 2&-1\\-1&2\end {array}
\right ]
 \een
is easily assigned the real pair of energies $E_\pm =2 \pm 1$ as
well as the general real-matrix ansatz for the metric
 \ben
 \Theta_0=\left [\begin {array}{cc} {f}&b\\b&{
 {f'}}\end {array}\right ]
 \een
reflecting its necessary Hermiticity. In an encouraging start of
our systematic study we may insert both these matrices in
eq.~(\ref{htot}) and arrive at the single constraint $f=f'$.

All the resulting two-parametric metrics $\Theta_0$ possess
eigenvalues $\theta_\pm$ expressible in closed form, $\theta_\pm =
f\pm b$. It is trivial to conclude that our $\Theta_0$ is positive
(and can be called a metric) iff $f=f'>0$ and $f^2>b^2$.

\subsection{All the metrics $\Theta_0$ at $N=4$.
\label{maaduo} }


In the first nontrivial step of our analysis let us consider the
four-dimensional Hamiltonian $H^{(4)}(\lambda)$ at $\lambda=0$,
 \ben
 H^{(4)}_0=H^{(4)}(0)=
 \left [\begin {array}{rrrr} 2&-1&0&0\\{}-1&2&-1&0
\\{}0&-1&2&-1\\{}0&0&-1&2\end {array}
\right ]\,
 \een
and let us try to deduce the generic form of all of the related
matrices $\Theta_0$ {\em directly} from the set of $N^2=16$
equations~(\ref{htot}) for the $N^2=16$ unknown (though,
presumably, real) matrix elements of $\Theta_0$. These equations
are not all linearly independent. No surprise -  the general
solution  contains $N=4$ real parameters \cite{Ali,SIGMA}.
%

\subsubsection{Construction \label{s1quo} }

As a typical task for Mathematica or Maple we solved our set by
the brute force methods of linear algebra and we obtained its
complete four-parametric solution
 \be
 \Theta^{(4)}_0=
\left [\begin {array}{cccc}
{\alpha_1}&{\alpha_2}&{\alpha_3}&{\alpha_4}
\\{}{\alpha_2}&{\alpha_1}+{\alpha_3}&
{\alpha_2}+{\alpha_4}& {\alpha_3}\\{}{\alpha_3}&{\alpha_2}+{
\alpha_4}&{\alpha_1}+{ \alpha_3}&{\it
\alpha_2}\\{}{\alpha_4}&{\alpha_3}&{ \alpha_2}&{ \alpha_1}\end
{array}\right ]
 \label{toposi}
 \ee
exhibiting linear dependence on all of its four parameters,
 \be
 \Theta^{(4)}_0=\alpha_1\,M_1+\alpha_2\,M_2+\alpha_3\,M_3+\alpha_4\,M_4\,.
 \label{superdzibe}
  \ee
While $M_1^{(4)}$ is just the four-dimensional unit matrix, the
remaining three expansion matrices represent its elementary
sparse-matrix generalizations,
 \ben %
 M_2^{(4)}=
 \left [\begin {array}{cccc}
  0&1&0&0\\{}
 1&0&1&0
 \\{}
 0&1&0&1\\{} 0&0&1&0\end {array}\right ]\,,\ \ \ \ \ \
 M_3^{(4)}=
 \left [\begin {array}{cccc}0&0&1&0\\{}0&1&0&1\\{}
  1&0&1&0
 \\{}
 0&1&0&0\end {array}\right ]\,,\ \ \ \ \ \
 M_4^{(4)}=
 \left [\begin {array}{cccc}
 0&0&0&1\\{} 0&0&1&0\\{}
 0&1&0&0\\{}
  1&0&0&0
  \end {array}\right ]\,.
 \een
This result of the computation indicates the possibility of the
existence of a certain friendly extrapolation pattern towards the
metrics $\Theta_0^{(N)}$ at any higher $N$.

\subsubsection{Positivity \label{s1quattro} }

After we specify the Hamiltonian but before we select any
particular solution $\Theta= \Theta(H)= \Theta^\dagger$ of
eq.~(\ref{tarov}) we have to guarantee that our candidate for the
metric is invertible and positive definite. Only then, this
operator can consistently specify the corresponding physical
Hilbert space ${\cal H}^{(physical)}$ of states of our quantum
system~\cite{Geyer}.

At the larger dimensions $N$ the proof of the positivity may be
difficult. In the four-dimensional matrix example (\ref{toposi})
it degenerates to the mere four elementary inequalities
 \ben
-2\,{\alpha_4}+2\,{\alpha_1}+{\alpha_3}-{ \alpha_2}\pm \sqrt {5}
\left (-{\alpha_2}+{ \alpha_3}\right )>0\,,
 \een
 \ben
 2\,{\alpha_4}+2\,{\alpha_1}+{\alpha_3}+{\it
\alpha_2}\pm \sqrt {5}\left ({\alpha_2}+{\alpha_3}\right )>0\,.
 \een
They must be satisfied as a guarantee of the positivity of all the
four eigenvalues $\theta_k$ of the metric $\Theta_0^{(4)}$.
%

\subsection{Extrapolation to
$N>4$}

\subsubsection{An ansatz for $\Theta^{(N)}_0$ \label{IVaauv} }

It is natural to expect  that formula (\ref{superdzibe}) is just the
first special case of the general expansion
 \be
 \Theta^{(N)}_0=\sum_{j=1}^N\, \alpha_j\,M_j^{(N)}(0)\,.
 \label{superdzi}
  \ee
Let us activate the experience collected at the smallest $N$ and
assume that all of the matrices $M=M_j^{(N)}(0)$ are solely
composed of the matrix elements 0 or 1. In the $j-$th matrix the
location of all of the non-vanishing elements may tentatively be
selected as follows,
 \be
 \left (M_j^{(N)}\right )_{ik}(0)= 1
 \ \ \ {\rm iff}\ \ \ i-k=m\,,\ \ \ N+1-i-k=n\,,\ \ \
 \ \ \ \ \ \ \
 \label{anza}
 \ee
 \ben
 \ \ \ \ \
 m=j-1, j-3, \ldots, 1-j\,,
 \ \ \ \ \ \ \ \
 n=N-j, N-j-2, \ldots, j-N\,.
 \een
Such an educated guess generalizes the above $N=2$ and $N=4$
results to all even dimensions. Its validity has carefully been
verified at several higher even integers  $N=2K$. One should note
that the mere insertion of the ansatz followed by the check of the
result is quick.

\subsubsection{Verification: $N=6$ etc.}

Formulae (\ref{superdzi}) and (\ref{anza}) determine all the
extrapolated $2K-$parametric matrices $\Theta^{(2K)}$. For
illustrative purposes let us pick up $N=2K=6$. This choice gives
the formula
 \ben
 \Theta^{(6)}_0=
\left [\begin {array}{cccccc} {\it \alpha_1}&{\it \alpha_2}&{\it
\alpha_3}&{\it \alpha_4}& {\it \alpha_5}&{\it \alpha_6}\\{}{\it
\alpha_2}&{\it \alpha_1}+{\it \alpha_3}&{ \it \alpha_2}+{\it
\alpha_4}&{\it \alpha_3}+{\it \alpha_5}&{\it \alpha_4}+{\it
\alpha_6}&{\it \alpha_5}
\\{}{\it \alpha_3}&{\it \alpha_2}+{\it \alpha_4}&{\it \alpha_1}+{\it \alpha_3}+
{\it \alpha_5}&{\it \alpha_2}+{\it \alpha_4}+{\it \alpha_6}&{\it
\alpha_3}+{\it \alpha_5}&{\it \alpha_4}
\\{}{\it \alpha_4}&{\it \alpha_3}+{\it \alpha_5}&{\it \alpha_2}+{\it \alpha_4}+
{\it \alpha_6}&{\it \alpha_1}+{\it \alpha_3}+{\it \alpha_5}&{\it
\alpha_2}+{\it \alpha_4}&{\it \alpha_3}
\\{}{\it \alpha_5}&{\it \alpha_4}+{\it \alpha_6}&{\it \alpha_3}+{\it \alpha_5}&
{\it \alpha_2}+{\it \alpha_4}&{\it \alpha_1}+{\it \alpha_3}&{\it
\alpha_2}\\{}{ \it \alpha_6}&{\it \alpha_5}&{\it \alpha_4}&{\it
\alpha_3}&{\it \alpha_2}&{\it \alpha_1}\end {array} \right ]\,.
 \een
We may easily verify the validity of the pertaining set of linear
equations (\ref{tarov}) by the simple-minded and straightforward
insertion again.

A different category of verification of the internal consistency
of our general $N-$parametric $N < \infty$ result (\ref{superdzi})
results from a direct study of the continuums limit $N \to
\infty$. A few comments on this interesting are collected in
Appendix D.

%

\section{Metrics $\Theta^{(N)}$ for $H=H^{(N)}(\lambda)$
at $\lambda > 0$ and $N \leq 6$\label{Vab}}

\subsection{Model with  $N=2$}

The key features of bound states in our discrete short-range
interaction models $H^{(N)}$  become already well illustrated via
their most elementary special case (\ref{dvojka}). Firstly, this
is the simplest model which shares peculiarity of the spectra
which remain real in the  $N-$independent interval of couplings
$\lambda \in (-1,1)$. For the whole sequence of our Hamiltonians
we shall parametrize $\lambda =\cos \varphi \in (-1,1)$,
therefore, with $\varphi \in (0,\pi/2)$.

At $N=2$ we can easily evaluate not only the closed formula for
the energies, $E=E^{(2)}_\pm =2\pm \sin \varphi$, but also the
norm $=2\,\sin^2\varphi$ of the related eigenstates $\psi_\pm$. In
addition, the cryptohermitian model $H^{(2)}(\cos \varphi )$
nicely illustrates the difference between its right eigenvectors
and their left-eigenstate partners. In the spirit and notation of
ref. \cite{SIGMA} these respective column vectors $|\psi_\pm
\rangle$ and their row-vector partners $\bbr \psi_\pm|$ are
different,
 \ben
 |\psi_\pm \rangle \sim \left (
 \ba
 1+\cos \varphi\\
 \mp \sin \varphi
 \ea
 \right )\,,
 \ \ \ \ \
 |\psi_\pm \kkt \sim \left (
 \ba
 1-\cos \varphi\\
 \mp \sin \varphi
 \ea
 \right )\,
 \een
but a biorthogonal basis can be formed of them. Thus,
$H^{(2)}(\cos \varphi )$ is a self-adjoint matrix in an {\em ad
hoc}, Hamiltonian-dependent Hilbert space of states ${\cal
H}^{(physical)}$.

In the light of refs. \cite{plb} a key merit of the $N=2$ example
can be seen in the existence of the explicit spectral definition
of the metric,
 \be
 \Theta=|\psi_+ \kkt\,t_+\,\bbr \psi_+|\,+
 |\psi_- \kkt\,t_-\,\bbr \psi_-|\,.
 \label{jednat}
 \ee
In this representation the guarantee of the necessary positivity
of the metric reads $t_\pm>0$. After the insertion of the
eigenvectors we arrive at another explicit formula for the metric,
 \be
 \Theta\, \sim\
 \left (
 \begin{array}{cc}
 (1-\cos \varphi)^2(t_++t_-)&
 (1-\cos \varphi)\sin \varphi(-t_++t_-)\\
 (1-\cos \varphi)\sin \varphi(-t_++t_-)&
 \sin^2 \varphi(t_++t_-)
 \ea
 \right )
 \,.
 \label{opera}
 \ee
Its inspection reveals that the metric may be re-written as a
superposition
 \be
 \Theta^{(2)}(\lambda)=\alpha_1\,M_1^{(2)}(\lambda)+\alpha_2\,M_2^{(2)}(\lambda)\,
 \ \ \ \ \ \ \ \ \ \ \ \ \ \
 \ee
with the two $\lambda-$dependent matrix coefficients,
 \be
 \ \ \ \ \ \ \ \ \ \ \ \
 M_1^{(2)}(\lambda)=
 \left [\begin {array}{cc} 1-\lambda &0\\{}0&1+\lambda\end {array}
\right ]\,, \ \ \ \ M_2^{(2)}(\lambda)=
 \left [\begin {array}{rr} 0&1\\{}1&0\end {array}
 \right ]\,.
 \label{super2}
  \ee
Such a re-parametrization leaves the positivity criterion entirely
transparent,
 \be
 \alpha_1>0, \ \ \ \ \ \alpha_1^2(1-\lambda^2)>\alpha_2^2\,, \ \ \ \ \ N=2\,
 \ee
so that we may choose any $\alpha_2$ from the interval
$(-\alpha_1\sin \varphi,\alpha_1\sin \varphi)$. The transition to
the Hermitian limit $\lambda \to 0$ appears facilitated in the new
parametrization.

\subsection{Shorthand notation \label{m2ttro} }

The continuity of the expansion matrices in the free-motion limit
as noticed above remains true at all $N>2$. Thus, we may visualize
metrics $\Theta^{(N)}(\lambda)$ as expanded in terms of the
generalized $\lambda-$dependent sparse matrix coefficients
obtained as a certain $\lambda-$deformation of their $\lambda = 0$
predecessors defined by closed formula~(\ref{anza}). With this
perspective in mind let us now define the following infinite
sequence of polynomials,
 \ben
 P_0=1\,,\ \ \
 P_1^{(\pm)}=1\pm \lambda\,,\ \ \
 P_2=1- \lambda^2\,,\ \ \
 P_3^{(\pm)}=(1\pm \lambda)\,(1- \lambda^2)\,,\ \ \
 \een
 \be
 P_4=(1- \lambda^2)^2\,,\ \ \
 P_5^{(\pm)}=(1\pm \lambda)\,(1- \lambda^2)^2\,,  \ \ \
 P_6=(1- \lambda^2)^3\,,\ \ \  \ldots\ .
 \ee
In terms of these polynomials the doublet of our sparse expansion
matrices $M_{1,2}^{(2)}(\lambda)$ can be characterized by the
``incidence" or ``indexing" matrices $S_{1,2}^{(2)}$ with certain
integer (or empty) entries. In general they will carry all the
information about the position and about the degree of polynomial
matrix elements $P_n=P_n(\lambda)$ of the respective matrices
$M_{j}^{(N)}(\lambda)$. At $N=2$ they are defined simply by the
following assignment,
 \ben
 M_1^{(2)}(\lambda)=
 \left [\begin {array}{cc} P_1^{(-)} &0\\{}0&P_1^{(+)} \end {array}
\right ]
 \ \ \ \
 \Longleftrightarrow
 \ \ \ \
 S_1^{(2)}=
 \left [\begin {array}{rr} 1&{\rm }\\{\rm }&1\end {array}
 \right ]\,,
 \een
 \ben
 M_2^{(2)}(\lambda)=
 \left [\begin {array}{cc} 0&P_0\\P_0&0 \end {array}
\right ]
 \ \ \ \
 \Longleftrightarrow
 \ \ \ \
 S_2^{(2)}=
 \left [\begin {array}{rr} {\rm }&0\\0&{\rm }
 \end {array}
 \right ]\,.
 \een
In what follows we shall demonstrate, step by step, that the
polynomials $P_{2k+1}^{(\pm)}(\lambda)$ with the minus-sign
superscript will always sit in the left upper triangle (i.e.,
above the second diagonal) of the respective expansion coefficient
$M_j^{(N)}(\lambda)$ and {\it vice versa}. Thus, at any $N$, the
expansion-coefficient functions $M_{j}^{(N)}(\lambda)$ will be
{\em unambiguously} determined by the mere indexing matrices
$S_{j}^{(N)}$. Of course, up to now we only demonstrated that
these observations are valid at $N=2$.

%

%
\begin{figure}[h]                     
\begin{center}                         
\epsfig{file=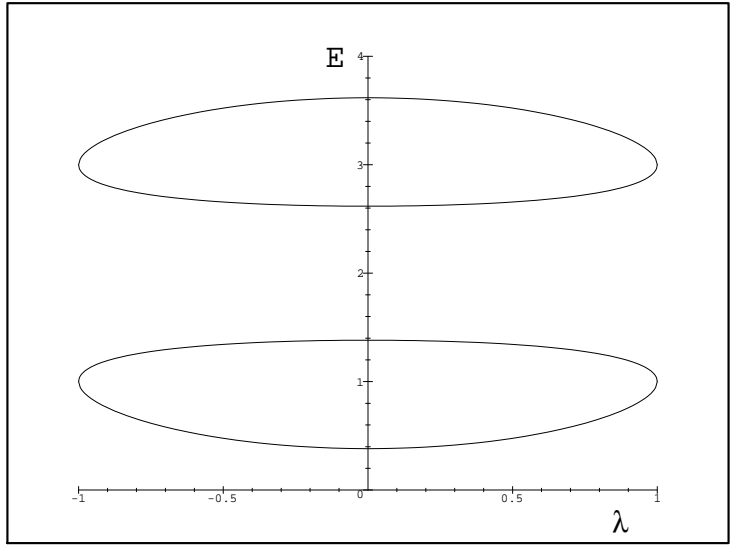,angle=270,width=0.7\textwidth}
\end{center}                         
\vspace{-2mm} \caption{Spectrum of $H^{(4)}(\lambda)$.
 \label{fione}}
\end{figure}

\subsection{Model with $N=4$ \label{mquattro} }

Hamiltonian  $H^{(N)}(\lambda)$ of eq.~(\ref{cetyrki}) is the
simplest model which is purely kinetic near its ``distant" lattice
points and which is dynamically nontrivial just in the vicinity of
the origin. The $\lambda-$dependent coupling merely connects the
two points $x_{N/2}$ and $x_{N/2+1}$ in the middle of the lattice.
The four eigenvalues of  matrix $H^{(4)}(\lambda)$ remain real in
the {\em same} interval of couplings $\lambda\in (-1,1)$ as above
(cf. Figure \ref{fione}),
 \ben
 E_{\pm,\pm}=2\pm \frac{1}{2}\,\sqrt {6-2\,{\lambda}^{2}
 \pm 2\,\sqrt {5-6\,{\lambda}^{2}+{\lambda}^{4}}}\,.
 \een
Symbolic manipulations on the computer enable us to find all the
corresponding matrices of the metric $\Theta^{(4)}(\lambda)$,
 \ben
 \left [\begin {array}{cccc} {\it \alpha_1}\,\left
(1-{\lambda}\right )&{\it \alpha_2}\, \left (1-{\lambda}\right
)&{\it \alpha_3}&{\it \alpha_4}\\\noalign{\medskip}{\it
\alpha_2}\, \left (1-{\lambda}\right )&{\it \alpha_1}\,\left
(1-{\lambda}\right )+{\it \alpha_3}\,\left (1-{\lambda} \right
)&{\it \alpha_4}+{\it \alpha_2}\,\left (1-{{\lambda}}^{2}\right
)&{\it \alpha_3}
\\\noalign{\medskip}{\it \alpha_3}&{\it \alpha_4}+{\it \alpha_2}\,\left (1-{{\lambda}}^{2}
\right )&{\it \alpha_1}\,\left (1+{\lambda}\right )+{\it
\alpha_3}\,\left (1+{\lambda}\right )&{ \it \alpha_2}\,\left
(1+{\lambda}\right )\\\noalign{\medskip}{\it \alpha_4}&{\it
\alpha_3}&{ \it \alpha_2}\,\left (1+{\lambda}\right )&{\it
\alpha_1}\,\left (1+{\lambda}\right )\end {array} \right ]\,.
 \een
They may again be interpreted as the sums
 \be
 \Theta^{(4)}(\lambda)=\alpha_1\,M_1+\alpha_2\,M_2+\alpha_3\,M_3+\alpha_4\,M_4\,
 \label{super4}
  \ee
where
 \ben
 M_1=
 \left [\begin {array}{cccc} 1-\lambda&0&0&0
 \\{}
 0&1-\lambda&0&0\\{} 0&0&1+\lambda&0\\{} 0&0&0&1+\lambda\end {array}\right ]\,,\ \ \ \ \ \
 M_2=
 \left [\begin {array}{cccc}
  0&1-\lambda&0&0\\{}
 1-\lambda&0&1-\lambda^2&0
 \\{}
 0&1-\lambda^2&0&1+\lambda\\{} 0&0&1+\lambda&0\end {array}\right ]\,,
 \een
 \be
 M_3=
 \left [\begin {array}{cccc}0&0&1&0\\{}0&1-\lambda&0&1\\{}
  1&0&1+\lambda&0
 \\{}
 0&1&0&0\end {array}\right ]\,,\ \ \ \ \ \
 M_4=
 \left [\begin {array}{cccc}
 0&0&0&1\\{} 0&0&1&0\\{}
 0&1&0&0\\{}
  1&0&0&0
  \end {array}\right ]\,.
  \label{sihi}
 \ee
In the shorthand notation of our previous paragraph the following
four  incidence matrices $S_j^{(4)}$ will carry again all the
necessary information about the respective four matrix polynomial
functions of $\lambda$. In computations, these incidence matrices
$S_j^{(4)}$, i.e.,
 \ben
 \left [\begin {array}{cccc} 1&{}&{}&{}
 \\{}
 {}&1&{}&{}\\{} {}&{}&1&{}\\{} {}&{}&{}&1\end {array}\right ]
 \,,\ \ \ \ \ \
 \left [\begin {array}{cccc}
  {}&1&{}&{}\\{}
 1&{}&2&{}
 \\{}
 {}&2&{}&1\\{} {}&{}&1&{}\end {array}\right ]
 \,,\ \ \ \ \ \
 \left [\begin {array}{cccc}{}&{}&0&{}\\{}{}&1&{}&0\\{}
  0&{}&1&{}
 \\{}
 {}&0&{}&{}\end {array}\right ]
 \,,\ \ \ \ \ \
 \left [\begin {array}{cccc}
 {}&{}&{}&0\\{} {}&{}&0&{}\\{}
 {}&0&{}&{}\\{}
  0&{}&{}&{}
  \end {array}\right ]\,
 \een
will be used for the encoding and/or efficient reconstruction of
the respective expansion matrices $M_j^{(4)}(\lambda)$.

It is worth noticing that even the simplest metric with
$\alpha_2=\alpha_3=\alpha_4$  which is proportional to the first
coefficient $M_1$ and which remains diagonal (i.e., in the
language of coordinates on the lattice, ``local") ceases to be
proportional to the unit matrix so that our model resides in a
nontrivial Hilbert space where $\Theta \neq I$.

%
\begin{figure}[h]                     
\begin{center}                         
\epsfig{file=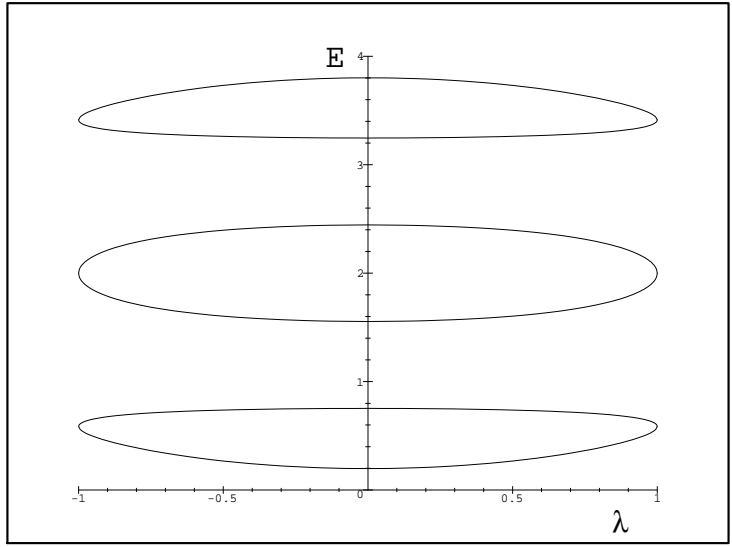,angle=270,width=0.7\textwidth}
\end{center}                         
\vspace{-2mm} \caption{Spectrum of $H^{(6)}(\lambda)$.
 \label{fitwo}}
\end{figure}

\subsection{Model with  $N=6$ \label{seisss} }

Although all the six eigenvalues of the matrix $H^{(6)}(\lambda)$
may be expressed in closed form in principle, we shall only
graphically confirm that all of them remain real in the same
interval as above, with $\lambda\in (-1,1)$ (cf. Figure
\ref{fitwo}). Inside this interval the metric
$\Theta^{(6)}(\lambda)$ exists and its general form is obtainable
from eq.~(\ref{tarov}) by its straightforward computer-assisted
solution. The resulting matrices $\Theta^{(6)}(\lambda)$ are
displayed here in the following split presentation,
 \ben
 \left [\begin {array}{cccc}
 {\it \alpha_1}\,\left (1-{\lambda}\right )&{\it \alpha_2}
\,\left (1-{\lambda}\right )&{\it \alpha_3}\,\left (1-{\lambda}\right )&\dots \\
{\it \alpha_2}\left (1-{\lambda}\right )&{\it \alpha_1} \left
(1-{\lambda}\right )+{\it \alpha_3}\left (1-{\lambda}\right )&{\it
\alpha_2}\,\left (1 -{\lambda}\right )+{\it \alpha_4}\,\left
(1-{\lambda}\right
)&\ldots\\
{ \it \alpha_3}\left (1-{\lambda}\right )&{\it \alpha_2}\left
(1-{\lambda}\right )+{\it \alpha_4} \left (1-{\lambda}\right
)&{\it \alpha_1}\left(1-{\lambda}\right )+{\it
\alpha_3}\left(1-{\lambda} \right )\left(1-{{\lambda}}^{2}\right
)+{\it \alpha_5}\left (1-{\lambda}\right )& \ldots
\\
{\it \alpha_4}&{\it \alpha_3}\,\left (1-{{\lambda}}^{2}\right )+{
\it \alpha_5}&{\it \alpha_2}\,\left (1-{{\lambda}}^{2}\right
)+{\it \alpha_4}\,\left (1-{{\lambda}}^{ 2}\right )+{\it
\alpha_6}&\ldots
\\
{\it \alpha_5}&{\it \alpha_4}+{\it \alpha_6}&{ \it
\alpha_3}\,\left (1-{{\lambda}}^{2}\right )+{\it \alpha_5}&\ldots
\\
{\it \alpha_6}&{\it \alpha_5}&{\it \alpha_4}&\ldots
\end {array}\right ]=
\een
 \ben
 =\left [\begin {array}{cccc}
 \ldots&{\it
\alpha_4}&{\it \alpha_5} &{\it \alpha_6}\\
\ldots&{\it \alpha_3}\,\left (1-{{\lambda}}^{2} \right )+{\it
\alpha_5}&{\it \alpha_4}+{\it \alpha_6}&{\it
\alpha_5}\\
\ldots&{\it \alpha_2}\,\left (1-{{\lambda}}^{2}\right )+{\it
\alpha_4}\,\left (1-{{\lambda}}^{2}\right )+{ \it \alpha_6}&{\it
\alpha_3}\,\left (1-{{\lambda}}^{2}\right )+{\it \alpha_5}&{\it
\alpha_4}
\\
\ldots&{\it \alpha_1}\left (1+{\lambda}\right )+{\it
\alpha_3}\left (1 +{\lambda}\right )\left (1-{{\lambda}}^{2}\right
)+{\it \alpha_5}\left (1+{\lambda}\right )&{\it \alpha_2}\left
(1+{\lambda}\right )+{\it \alpha_4}\left (1+{\lambda}\right )&{\it
\alpha_3} \left (1+{\lambda}\right
)\\
\ldots&{\it \alpha_2}\left (1+{\lambda} \right )+{\it
\alpha_4}\left (1+{\lambda}\right )&{\it \alpha_1}\left
(1+{\lambda}\right )+{ \it \alpha_3}\left (1+{\lambda}\right
)&{\it \alpha_2}\left (1+{\lambda}\right )
\\
\ldots&{\it \alpha_3}\,\left (1+ {\lambda}\right )&{\it
\alpha_2}\,\left (1+{\lambda}\right )&{\it \alpha_1}\,\left
(1+{\lambda}\right )
\end {array}\right ].
 \een
The use of the shorthand symbols $S_j^{(6)}$ becomes indispensable
for the sufficiently efficient and compact encoding of the $N=6$
expansion formula
 \be
 \Theta^{(N)}(\lambda)=\sum_{j=1}^N\,
 \alpha_j\,M_j^{(N)}(\lambda)\,.
 \label{superdzika}
  \ee
At $j=1,2,\ldots,6$ the six respective compact incidence or
indexing matrices ${S}_j^{(6)}$ are
 \ben
 \left [\begin {array}{cccccc} 1&{}&{}&{}&{}&{}
 \\{}
 {}&1&{}&{}&{}&{}\\{} {}&{}&1&{}&{}&{}\\{} {}&{}&{}&1&{}&{}
 \\{} {}&{}&{}&{}&1&{}
 \\{} {}&{}&{}&{}&{}&1
 \end {array}\right ]\,,\ \ \ \ \ \
 \left [\begin {array}{cccccc}{}&1&{}&{}&{}&{}\\{}  1&{}&1&{}&{}&{}
 \\{}
 {}&1&{}&2&{}&{}\\{} {}&{}&2&{}&1&{}\\{} {}&{}&{}&1&{}&1
 \\{} {}&{}&{}&{}&1&{}
 \end {array}\right ]\,,\ \ \ \ \ \
 \left [\begin {array}{cccccc} {}&{}&1&{}&{}&{}\\{} {}&1&{}&2&{}&{}
 \\{} 1&{}&3&{}&2&{}
 \\{} {}&2&{}&3&{}&1
 \\{}
 {}&{}&2&{}&1&{}\\{} {}&{}&{}&1&{}&{}
 \end {array}\right ]\,,\ \ \ \ \ \
 \een
 \ben
 \left [\begin {array}{cccccc} {}&{}&{}&0&{}&{}
 \\{}
 {}&{}&1&{}&0&{}\\{} {}&1&{}&2&{}&0\\{} 0&{}&2&{}&1&{}
 \\{} {}&0&{}&1&{}&{}
 \\{} {}&{}&0&{}&{}&{}
 \end {array}\right ]\,,\ \ \ \ \ \
 \left [\begin {array}{cccccc} {}&{}&{}&{}&0&{}
 \\{}
 {}&{}&{}&0&{}&0\\{} {}&{}&1&{}&0&{}\\{} {}&0&{}&1&{}&{}
 \\{} 0&{}&0&{}&{}&{}
 \\{} {}&0&{}&{}&{}&{}
 \end {array}\right ]\,,\ \ \ \ \ \
 \left [\begin {array}{cccccc} {}&{}&{}&{}&{}&0
 \\{}
 {}&{}&{}&{}&0&{}\\{} {}&{}&{}&0&{}&{}\\{} {}&{}&0&{}&{}&{}
 \\{} {}&0&{}&{}&{}&{}
 \\{} 0&{}&{}&{}&{}&{}
 \end {array}\right ].
 \een
These matrices form an inseparable part of any practical
computer-assisted application of the formalism at $N\geq 4$. At
$N=6$, moreover, their explicit form also offers an interesting
insight in their $N-$dependence forming a sufficiently inspiring
starting point of our extrapolation programme.

\section{Extrapolation \label{VIab}}

Starting from $N = 8$, the explicit form of the indexing matrices
$S_j^{(N)}$ becomes also rather large for being printed. Still,
their computer-assisted use remains as easy and straightforward as
at $N = 10$ etc. Thus, in the main part of our homework we simply
formulated and tested the alternative  extrapolation hypotheses.
Now, it remains for us just to summarize the results.

\subsection{Recurrences for the incidence matrices, with the central ones exempted.}

It is easy to reconstruct the $\lambda-$dependent matrices
$M_j^{(2K)}(\lambda)$ from the knowledge of the respective
shorthand symbols $S_j^{(2K)}$. What remains for us to do is to
define all the set of the shorthand symbols ${S}_j^{(2K)}$
representing all our expansion matrices. In such a context, the
trial and error method enabled us to collect a sufficiently
extensive set of these symbols. In the next step, their inspection
revealed that  at any given $N=2K$ the first $K-1$ matrices
${S}_j^{(2K)}$ with $j=1,2,\ldots,K-1$ would be easily constructed
from their predecessors ${S}_j^{(2K-2)}$.

Purely empirically, the latter recurrent construction has been
found in an enlargement of the dimension followed by a symmetric
attachment of the two $j-$plets of units $``1"$ in the empty parts
of the left upper corner and of the right lower corner.

In an entirely similar manner, the last $K$ matrices
${S}_{2K+1-j}^{(2K)}$ with $j=1,2,\ldots,K$ become also formed in
the similar manner. Explicitly, their $K$ predecessors
${S}_{2K-1-j}^{(2K-2)}$ must be modified by attaching $j$ zeros
$``0"$ in the right upper corner and in the left lower corner.

In both of the ``leftmost-subsequence" and
``rightmost-subsequence" scenarios,  the results displayed in
section~\ref{Vab} offer a sufficiently instructive illustration of
such a recipe. An explicit algebraic reformulation of such a
doublet of two-dimensional recurrences would be also as
straightforward and compact as their above, purely verbal
description. An algorithmic version of these recurrences was,
after all, needed also during our practical computer-assisted
evaluation of the matrix elements of $\Theta^{(N)}(\lambda)$ at
all the $N$ which we considered. Of course, all these descriptions
of our two-dimensional recurrences are equivalent and amenable to
the rigorous proof by mathematical induction. Due to the lack of
space, the routine details of algebra as well as of its rigorous
proofs are skipped here and left to interested readers.

\subsection{Recurrences for the central incidence matrices ${S}_K^{(2K)}$ }

At the remaining subscript $j=K$ the construction of the most
complicated missing member ${S}_K^{(2K)}$ of the family  must be
discussed separately. In both the ``leftmost-subsequence" and
``rightmost-subsequence" scenarios its $(2K-2)$ by $(2K-2)$
predecessor  proves, rather unexpectedly, {\em different} from the
naively expected matrix ${S}_K^{(2K-2)}$.

This means that the sequence of the ``middle" or ``central"
matrices ${S}_K^{(2K)}$ should be treated as exceptional.
Fortunately, they remain created by a straightforward recurrent
recipe. Its idea relies on the use of certain specific predecessor
matrices ${\cal L}^{(2K-2)}$. With a freedom in their
specification let us decide to proceed in the
``rightmost-subsequence" manner. This means we shall enlarge the
dimension of ${\cal L}$ () whatever it is) and we shall fill $K$
``neighboring" units $``1"$ in the left upper corner and in the
right lower corner.

We are now ready to define the specific predecessors ${\cal
L}^{(2K-2)}$. They appear to be constructed from the old ``middle"
matrices ${S}_{K-1}^{(2K-2)}$ via a specific two-step recipe.
Firstly we replace each ``old" numerical element in
${S}_{K-1}^{(2K-2)}$ by its successor, i.e., we replace ``old 0"
by ``1", ``old 1" by ``2", etc. In the second step we form a
left-right reflection of the resulting matrix and arrive at the
final form of the necessary predecessor ${\cal L}^{(2K-2)}$ as a
result. Thus, at $N=4$ we have the sequence
 \ben
  {S}_1^{(2)}=
 \left [\begin {array}{cc} 1 &{\rm }\\{}{\rm }&1\end {array}
\right ]\ \to \
 \left [\begin {array}{cc} 2 &{\rm }\\{}{\rm }&2\end {array}
\right ]\ \to \ {\cal L}^{(2)}=
 \left [\begin {array}{cc}  &2\\{}2&{}\end {array}
\right ]\ \to \
  {S}_2^{(4)}\,.
 \een
Similarly, the recurrent construction of the matrix
${S}_{N/2}^{(N)}$ at $N=6$  will result from adding six units ``1"
to the auxiliary predecessor matrix ${\cal L}^{(4)}$ in the
formula
 \ben
  {S}_2^{(4)}\ \to \
 \left [\begin {array}{cccc}
  {}&2&{}&{}\\{}
 2&{}&3&{}
 \\{}
 {}&3&{}&2\\{} {}&{}&2&{}\end {array}\right ]\ \to \ {\cal L}^{(4)}=
 \left [\begin {array}{cccc}
  {}&{}&2&{}\\{}
 {}&3&{}&2
 \\{}
 2&{}&3&{}\\{} {}&2&{}&{}\end {array}\right ]
 \ \to \
  {S}_3^{(6)}\,
 \een
etc.  We may conclude that the central matrices ${S}_K^{(2K)}$ at
the respective $K=1,2,3,4$ (etc) form the sequence
 \ben
 \left [\begin {array}{cc} 1 &\\&1\end {array}
 \right ], \ \ \ \
 \left [\begin {array}{cccc}
  &1&&\\{}
 1&&2&
 \\{}
 &2&&1\\{} &&1&\end {array}\right ],\ \ \ \
 \left [\begin {array}{cccccc} &&1&&&\\{} &1&&2&&
 \\{} 1&&3&&2&
 \\{} &2&&3&&1
 \\{}
 &&2&&1&\\{} &&&1&&
 \end {array}\right ],\ \ \ \
 \left [\begin {array}{cccccccc} &&&1&&
 &&\\
 &&1&&
 {2}&&&\\
 &1 &&3&&{2}
 &&\\
 1&&3&&4&&{2}
 &\\
 &{2}
 &&4&&3&&1
 \\
 &&{2}&&3&&
 1 &\\
 &&&
 {2}&&1
 &&\\
 &&&&1&&&\end
 {array}\right ]
 \een
(etc). The general pattern of their recurrent construction is
obvious.

\subsection{Verifications}

The full-fledged formulae for the eight-parametric
$\Theta^{(8)}(\lambda)$ already cease to be easily printable but
their characterization using the incidence matrices remains fully
transparent and compact. All of the individual expansion matrices
entering the general series~(\ref{superdzika}) for metrics
$\Theta$ exhibit the same simultaneous change of the sign of
${\lambda}$ after the reflection with respect to their second
diagonal. This is well visible in our last illustrative equation
 \ben
 \left [\begin {array}{cccccc} &&&1-{\lambda}&&
 \\
 &&1-{\lambda}&&
 1-{{\lambda}}^{2}&\\
 &1 -{\lambda}&&\left (1-{\lambda}\right )\left
 (1-{{\lambda}}^{2}\right )&&\ldots
 \\
 1-{\lambda}&&\left (1-{\lambda}\right )\left (1-{{\lambda}}^{2} \right )&&\left
 (1-{{\lambda}}^{2}\right )^{2}&\ldots
 \\
 &1-{{\lambda}}^{2}
 &&\left (1-{{\lambda}}^{2}\right )^{2}&& \ldots
 \\
 &&1-{{\lambda}}^{2}&&\left (1+{\lambda}\right )\left (1-{{\lambda}}^{2}\right )& \ldots\\
 &&&
 1-{{\lambda}}^{2}&&\ldots
 \\
 &&&&1+{\lambda}&\end{array}\right ]
 \een
where a part of the real and symmetric matrix
$M_K^{(2K)}(\lambda)$ at $K=4$ is displayed.

\section{Summary \label{VI} \label{summary} }


In quantum theory an operator $H$ represents an observable
provided only that it is self-adjoint in a Hilbert space equipped
with a metric $\Theta$.  For a given $H$, equation $H^\dagger
\Theta = \Theta H$ specifies a complete menu of all the eligible
$\Theta=\Theta(H)$ needed to determine the inner product. We
illustrated the feasibility of the construction of all of these
$\Theta$s for an infinite sequence of certain one-parametric $N$
by $N$ matrices $H^{(N)}(\lambda)$ with $\lambda \in (-1,1)$.

A recurrent method of the construction of all the admissible
metric matrices $\Theta^{(N)}(\lambda)$  has been proposed and
tested at $\lambda=0$. For  $\lambda \in (-1,1)$, the
straightforward construction of the individual $N-$parametric
$\Theta^{(N)}(\lambda)=
\Theta^{(N)}_{\alpha_1,\alpha_2,\ldots,\alpha_N}(\lambda)$ was
based on the computer-assisted symbolic manipulations at the
smallest $N\leq N_{(minimal)}$. Next we extrapolated these
formulae for the metric to all the subsequent larger
$N>N_{(minimal)}$. We carefully verified the validity of our
extrapolations via tests performed at a number of sample
$N>N_{(minimal)}$.

We revealed that the recurrences which are needed for the
reconstruction of all the set of all the $N-$parametric  matrices
$\Theta^{(N)}(\lambda)=
\Theta^{(N)}_{\alpha_1,\alpha_2,\ldots,\alpha_N}(\lambda)$ from
the set of their $N-1$ predecessors
 $\Theta^{(N-1)}(\lambda)=
\Theta^{(N-1)}_{\alpha_1,\alpha_2,\ldots,\alpha_{N-1}}(\lambda)$
degenerate to the recurrences needed for the reconstruction of the
$N-$plets of matrix coefficients $M^{(N)}_j(\lambda)$. The
ultimate simplification of our recurrent recipe has been achieved
when all the matrices $M^{(N)}_j(\lambda)$ proved easily
constructed from the knowledge of the related elementary indexing
matrices $S^{(N)}_j$ with integer or empty entries.

During our study of our matrix toy-model Hamiltonians we persuaded
ourselves that the underlying mathematics is friendly and that the
recurrences generate the matrix indices $S^{(N)}_j$ in a really
transparent form. Although this fact should partially be
attributed to the mere one-parametric choice of our class of
Hamiltonians $H^{(N)}(\lambda)$, we firmly believe that our method
which mixed the low$-N$ evaluations with subsequent all$-N$
extrapolations might keep its efficiency for a number of more
realistic and, in particular, more-parametric sparse-matrix
toy-model Hamiltonians.

%
%
%

\vspace{15mm}

\section*{Acknowledgement}

Work supported by the M\v{S}MT ``Doppler Institute" project Nr.
LC06002,  by the Institutional Research Plan AV0Z10480505 and by
the GA\v{C}R grant Nr. 202/07/1307.


\section*{Figure captions}

\subsection*{Figure 1. Spectrum of $H^{(4)}(\lambda)$.}

\subsection*{Figure 2. Spectrum of $H^{(6)}(\lambda)$.}

\vspace{5mm}

\newpage

\section*{Appendix A: Toy-model potentials in eq.~(\ref{basic}) }

Among the oldest explicit examples of the phenomenological
Hamiltonian operators $H$ exhibiting the cryptohermiticity
property (\ref{newh}) one may cite, e.g., the imaginary cubic
anharmonic oscillators studied by Caliceti et al \cite{Caliceti},
the ``wrong-sign" quartic anharmonic oscillators described by
Buslaev and Grecchi and others \cite{BG} or a family of the more
general, ``non-Hermitian" but stable bound-state models as
proposed by Bender and Boettcher \cite{BB}. In all of these
one-dimensional schematic models of bound states the
reconstruction of {\em any} suitable metric $\Theta$ from a given
oscillator Hamiltonian $H$ proves rather difficult. {\it Pars pro
toto} we may recall the perturbation-series construction of
$\Theta(H)$ for the imaginary cubic $H$ \cite{cubic}, with many
further relevant references cited therein.

A perceivably better picture of the properties of the
physics-determining metrics $\Theta(H)$ has been obtained for
certain exactly solvable potentials of bound states (cf. the
semi-numerical construction of $\Theta(H)$ for a square-well model
$H\neq H^\dagger$ \cite{Batal} as an example).

The situation appeared perceivably worsened after transition to
the scattering regime where even the combination of a
sophisticated perturbation construction with the choice of the
really most elementary exactly solvable interaction potentials in
eq.~(\ref{basic}) did leave many conceptual questions unanswered
(cf., e.g., refs. \cite{Jones,delta}).


\section*{Appendix B:
Conventional choices of $\Theta=\Theta(H)$ in quantum theory
 }

In the majority of applications of quantum theory using non-Dirac
Hilbert spaces ${\cal H}^{(physical)}$ the construction of the
metric remains almost prohibitively difficult even when one
restricts attention to the {\em single} observable, i.e., to $j=0$
in eq.~(\ref{newhtot}). One may feel forced to work just with a
drastically simplified and/or highly schematic class, say, of
point-interaction Hamiltonians \cite{delta,Krejcirik}. The
construction of $\Theta(H)$ may remain feasible only when one
employs an approximate (e.g., perturbation \cite{cubic,Batal})
method. Still, even in many papers which define a quantized system
in an unusual Hilbert space ${\cal H}^{(physical)}$ which is
equipped with a non-Dirac metric $\Theta \neq I$ their authors
usually avoid the overcomplicated requirement (\ref{newhtot}) and
select just one of {\em particular } solutions $\Theta(H)$ of
eq.~(\ref{htot}) instead.

As one of the the simplest illustrations of such a strategy we
might recall even the most common model (\ref{basicss}) with a
most common real potential. Evidently,  just the trivial solution
$\Theta^{(Dirac)}=I$ of eq.~(\ref{htot}) is being assigned to
$H=H^\dagger$. In particular, out of all of the above-cited
eligible metrics $\Theta_0^{(Mostafazadeh)}(F,K)$ just the most
trivial particular solution with $F=K=0$ is being used in
connection with the free-motion version of eq.~(\ref{basic}).

In  ${\cal PT}-$symmetric quantum mechanics admitting $\Theta\neq
\Theta^{(Dirac)}$  (cf. its thorough recent review written by Carl
Bender \cite{Carl}), the problem of the  ambiguity of the choice
of the metric $\Theta=\Theta(H)$ has been circumvented as well.
Although this less traditional formalism admits various
nonstandard, apparently non-Hermitian models (including even field
models with real spectra  \cite{Fring} etc), the current choice of
the space ${\cal H}^{(physical)}$ is equally restrictive,
preferring special metrics $\Theta^{(Bender)}={\cal CP}$ where
${\cal C}={\cal C}(H)$ represents a unique ``charge" while ${\cal
P}$ is the usual parity.

The most natural generalization of the ${\cal PT}-$symmetric
theories with $\Theta\neq \Theta^{(Dirac)}$ has been described by
Mostafazadeh \cite{Ali}. He re-attracted the attention of the
international scientific community to the abstract quantization
rule (\ref{newhtot}) of ref.~\cite{Geyer} and to the related
ambiguity of the reconstruction of the correct Hilbert space
${\cal H}^{(physical)}$ form a given Hamiltonian. He worked out
some illustrative examples (cf. \cite{cubic}) and, together with
Batal \cite{Batal}, he emphasized the possible physical relevance
of metrics $\Theta \neq \Theta^{(Bender)}$.


\section*{Appendix C: Discretized Runge-Kutta version of
eq.~(\ref{basic}) }

One of the key simplifications of some of the technical aspects of
solving differential Schr\"{o}dinger eq.~(\ref{basic}) is commonly
sought in its replacement by its Runge-Kutta difference-equation
approximation
 \be
  -
 \frac{\psi(x_{k+1})-2\,\psi(x_{k})+\psi(x_{k-1})}{h^2}
  +V(x_k)
 \,\psi(x_{k})
 =E\,\psi(x_k)\,
 \label{diskr}
 \ee
(cf., e.g., ref. \cite{thatwork} for more details). In place of
the real line of coordinates $x \in I\!\!R$ the equidistant
lattice of points $x_k$ may be conveniently defined by the formula
 $
  x_k=-1+kh$, $k \in Z\!\!\!\!Z$
in terms of a suitable (i.e., usually, sufficiently small)  real
constant $h>0$. For the purposes of the description of scattering
the lattice remains infinite so that the kinetic energy operator
$-d^2/dx^2$ may be visualized as the following tridiagonal matrix
 \be
 H_0=
 \left [\begin {array}{rrrc|crrr}
  \ddots&\ddots&&&&&&
  \\{}
  \ddots&2&-1&&&&&
 \\{}
 &-1&2&-1&&&&
 \\{}
 &&-1&2&-1&&&
 \\
  \hline
 &&&  -1&2&-1&&
 \\{}
 &&&&-1&2&-1&
 \\{}
 &&&&&-1&2&\ddots
 \\{}
 &&& &&&\ddots&\ddots
 \end {array}\right ]\,.
 \label{tridif0}
 \ee
In the bound-state context with the Dirichlet boundary conditions
$\psi(-1)=\psi(1)=0$ one usually considers just the finite set of
the lattice points,
 $k =  1, 2, \ldots, N$. With $x_{N+\,1}=+1$
we, in effect, fix an elementary length $
  h = {2}/({N+1})$ which would vanish in the continuum limit $N \to
\infty$.

In the latter scenario the explicit specification (\ref{newh}) of
the non-Dirac Hilbert space ${\cal H}^{(physical)}$ must be
slightly modified,
 \be
 ( \psi,\psi')_\Theta = \sum_{n,n'\in Z\!\!\!Z}\,\psi^*(x_n)\,
 \Theta(x_n,x_{n'})\,\psi'(x'_{n'})\,,\ \ \ \ \ \
   H = H^\ddagger\,.
 \label{newhdi}
 \ee
On this background the numerical use of the approximation
(\ref{basic}) $\longrightarrow$ (\ref{diskr}) finds numerous
applications in cryptohermitian quantum mechanics. In papers
\cite{jedna}, for example, the exact, analytic solvability of the
differential eq.~(\ref{basic}) with a certain class of
sufficiently simple complex potentials $V(x)\neq V^*(x)$ has been
shown {paralleled} by the exact solvability of the discrete
partner eq.~(\ref{diskr}).

Whenever the potential $V(x)$ remains sufficiently smooth, the
role of the discretization errors may be expected negligible, in
the domain of the sufficiently large $N \gg 1$ at least.
Difficulties may arise for point interactions. They proved popular
\cite{Jakubsky} and  found  applications in relativistic equations
\cite{jakg} and in manybody systems \cite{Fei}. This, as we
already mentioned, motivated our interest in in the new field of
cryptounitary scattering \cite{jpaft} and, in particular, in its
simplest discrete model~(\ref{tridif}).


\section*{Appendix D: Models $H^{(N)}(\lambda)$ at large $N \gg 1$ }

At a sufficiently large $N$ our one-parametric Hamiltonian
$H^{(N)}(\lambda)$ represents in fact one of the discrete versions
of eq.~(\ref{basic}) with a certain point interaction $V(x)$
localized in the origin. For a deeper understanding of such a
correspondence let us asccept that $x_0=-1$ and $x_{2K+1}=1$ and
let us abbreviate $2-h^2E=\cos \epsilon$ as usual
\cite{jpaft,jedna}. We may then treat $\epsilon\in (0,\pi)$ as a
new energy variable and we may visualize the wave functions
$\psi(x)$ with $x \neq 0$ as satisfying the free-motion eq.
(\ref{diskr}), complemented by the boundary conditions
$\psi(x_0)=\psi(x_{2K+1})=0$. This picture of the $N\gg 1$ system
must be completed by the doublet of the $\lambda-$dependent
relations
 \be
 (1+\lambda)\,{\psi(x_{{K}+1})-2\cos
 \epsilon\,\psi(x_{{K}})+\psi(x_{{K}-1})}=0\,,
 \label{diskra}
 \ee
 \be
 {\psi(x_{{K+2}})-2\cos
 \epsilon\,\psi(x_{{K+1}})+(1-\lambda)\, \psi(x_{{K}})}=0\,
 \label{diskrb}
 \ee
so that we may assume the emergence of a discontinuity of
$\psi(x)$ in the origin.

At the sufficiently large $N \sim 1/h \gg 1$ the wave functions
near $x=0$ remain well represented by their respective one-sided
Taylor series so that eqs.~(\ref{diskra}) and (\ref{diskrb}) may
be interpreted simply as a matching condition. We return to the
original energy variable $h^2E=2-2\cos \epsilon\equiv F$ and
insert the truncated expansions
 \ben
 \psi(x_{{K}-1})=\psi_L(0) -\frac{3}{2}\,h\,\psi_L'(0)
  +{\cal O}(h^2)\,,\ \ \ \ \
 \psi(x_{{K}})=\psi_L(0) -\frac{1}{2}\,h\,\psi_L'(0)
  +{\cal O}(h^2)\,,
 \een
 \ben
 \psi(x_{{K}+1})=\psi_R(0) +\frac{1}{2}\,h\,\psi_R'(0)
  +{\cal O}(h^2)\,,\ \ \ \ \
 \psi(x_{{K}+2})=\psi_R(0) +\frac{3}{2}\,h\,\psi_R'(0)
  +{\cal O}(h^2)\,
 \een
in eqs.~(\ref{diskra}) and (\ref{diskrb}). A straightforward
algebra leads to the following elementary condition
 \be
 \frac{h}{2}\,
 \left (
 \begin{array}{cc}
 -(1+\lambda)&F+1\\
 -(F+1)&1-\lambda
 \ea
 \right )\,
 \left (
 \ba
 \psi_R'(0)\\
 \psi_L'(0)
 \ea
 \right )=
 \left (
 \begin{array}{cc}
 1+\lambda&F-1\\
 F-1&1-\lambda
 \ea
 \right )\,
 \left (
 \ba
 \psi_R(0)\\
 \psi_L(0)
 \ea
 \right )\,
 \label{mbc}
 \ee
which matches the wave functions and their derivatives in the
origin. In the domain of sufficiently small $h>0$ this relation is
equivalent to the original constraints (\ref{diskra}) and
(\ref{diskrb}). We may conclude that  in the continuum limit $N\to
\infty$ our sequence of the matrix Hamiltonians $H^{(N)}(\lambda)$
can be reinterpreted as a series of dynamical models which
converge to a  specific differential eq.~(\ref{basic}) which is
split in two halves. Indeed, the $h\to 0$ limit of eq.~(\ref{mbc})
produces the elementary opaque-wall constraint
$\psi_R(0)=\psi_L(0)=0$.

At all the  nonvanishing small ``elementary lengths" $h>0$, our
rigorous definition (\ref{mbc}) leaves the point-interaction
potential term $V(x)$ translucent and manifestly energy-dependent.
Its definition by the mixed boundary conditions is nonlinear in
the coupling $\lambda$. Various special cases of this $N\gg 1$
bound-state model may be studied noticing, for example, that the
energy-dependence disappears in the low-excitation regime where
the quantity $F=h^2E$ remains negligible.

Another interesting special case is encountered when the
interaction is completely switched off, $\lambda \to 0$. Then we
may recollect our $N-$parametric free-motion formula
(\ref{superdzi}) for the metric $\Theta_0$ and we may compare it
with its two-parametric differential-operator counterpart
(\ref{most}). This comparison confirms that a linear combination
of the matrices $M_1^{(N)}(0)$ and $M_N^{(N)}(0)$ survives the
limiting transition $N\to \infty$ while, in contrast, all the
other terms in (\ref{superdzi}) disappear due to the build-in
requirement of the absence of an elementary length in the
Mostafazadeh's theory of ref.~\cite{cubic}.

\end{document}